\documentclass[11pt]{article}
\usepackage{jcapmod}

\usepackage{booktabs}
\usepackage[english]{babel}
\usepackage{amsmath,amssymb,amsbsy,amstext, amsthm, simplewick}
\usepackage{hyperref}
\usepackage{graphicx}
\usepackage{amsfonts}
\usepackage{amssymb}
\usepackage{upgreek}
\usepackage{simplewick}
 \usepackage{exscale,relsize}

\usepackage[margin=1cm,labelfont={sf,bf,scriptsize},textfont={sf,scriptsize}]{caption}

\usepackage{colortbl}
\definecolor{lightgreen}{cmyk}{0.2, 0, 0.2, 0.2}
\definecolor{lightgray}{cmyk}{0.1,0.2,0,0.1}
\definecolor{lightgray2}{cmyk}{0.1,0.1,0,0.1}

\makeatletter
\newlength{\apb@width}
\newcommand{\autoparbox}[2][c]{\settowidth{\apb@width}{#2}\parbox[#1]{\apb@width}{#2}}

\makeatother

\numberwithin{equation}{section}

\def\beq{\begin{equation}}
\def\eeq{\end{equation}}

\def\bea{\begin{eqnarray}}
\def\eea{\end{eqnarray}}

\def\beq{\begin{equation}}
\def\eeq{\end{equation}}
\def\bea{\begin{eqnarray}}
\def\eea{\end{eqnarray}}
\def\epsosc{\epsilon_{\rm osc}}

\def\R{\zeta}
\def\Mp{M_{\rm pl}}

\def\H{{\cal H}}
\def\L{{\cal L}}

\def\0{{\vec{0}}}
\def\k{{\vec{k}}}

\def\x{{\vec{x}}}

\def\p{{\vec{p}}}

\DeclareRobustCommand{\SkipTocEntry}[4]{}

\def\Mp{M_{\rm pl}}

\newcommand{\mbf}[1]{\mathbf #1}

\newcommand{\Expect}[1]{\left\langle #1 \right\rangle}
\def\H{{\rm H}}

\DeclareSymbolFont{extraup}{U}{zavm}{m}{n}
\DeclareMathSymbol{\varheart}{\mathalpha}{extraup}{86}
\DeclareMathSymbol{\vardiamond}{\mathalpha}{extraup}{87}

\begin{document}

\begin{titlepage}

\setcounter{page}{1} \baselineskip=15.5pt \thispagestyle{empty}

\bigskip\

\vspace{2cm}
\begin{center}
{\fontsize{19}{36}\selectfont  \sc Collective Symmetry Breaking and \\ \vskip 10pt Resonant Non-Gaussianity}
\end{center}

\vspace{0.6cm}

\begin{center}
{\fontsize{13}{30}\selectfont  Siavosh R.~Behbahani$^{\spadesuit}$ and Daniel Green$^\clubsuit$}
\end{center}

%\vspace{0.2cm}

\begin{center}
\vskip 8pt
\textsl{$^\spadesuit$  Physics Department, Boston University, Boston, MA 02215}

\vskip 7pt
\textsl{$^\clubsuit$ School of Natural Sciences,
 Institute for Advanced Study,
Princeton, NJ 08540, USA}

\end{center}

\vspace{1.2cm}
\hrule \vspace{0.3cm}
{ \noindent \textbf{Abstract} \\[0.2cm]
\noindent 
We study inflationary models that produce a nearly scale-invariant power spectrum while breaking scale invariance significantly in the bispectrum.  Under most circumstances, such models are finely-tuned, as radiative corrections generically induce a larger signal in the power spectrum.  However, when scale invariance is broken collectively (i.e., it requires more than one coupling to break the symmetry), these radiative corrections may be suppressed.  We illustrate the features and limitations of collective symmetry breaking in the context of resonant non-gaussianity. We discuss two examples where oscillatory features can arise predominantly in the bispectrum.}

 \vspace{0.3cm}
 \hrule

\vspace{0.6cm}
\end{titlepage}

\tableofcontents

\newpage
\section{Introduction}

One of the remarkable features of inflationary cosmology is the robustness of the (nearly) scale invariant power spectrum.  Despite a vast number of mechanisms that give rise to inflation, their predictions for the power spectrum typically differ at the percent level.  By contrast, higher $N$-point correlation functions can have vastly different behavior and may allow us to distinguish these mechanisms.

The origin of this similarity is an approximate time translation symmetry of the effective Lagrangian for the fluctuations.  Scale invariance arises in the presence of a time translation symmetry because each mode experiences the same history and freezes with nearly the same amplitude.   This symmetry arises from a shift symmetry for the inflaton, $\phi \to \phi + c$, which implies the theory is insensitive to the location of the background field on its potential.

Shift symmetries in models of inflation have a long history.  In the context of slow-roll inflation, it is well known that an approximate shift symmetry is required to protect the small mass of the inflaton from radiative corrections and from dangerous irrelevant operators (see e.g. \cite{Linde:1983gd, Freese:1990rb}).  Additional work \cite{Kaplan:2003aj, ArkaniHamed:2003mz, Silverstein:2008sg, McAllister:2008hb,Berg:2009tg,Baumann:2010ys, Baumann:2010nu} has provided explanations for such a symmetry from ultra-violet (UV) physics, yielding concrete realizations of these older ideas.  

Recently, it has been suggested that only a discrete shift symmetry is required for viable models of inflation \cite{Flauger:2009ab, Flauger:2010ja, Barnaby:2011qe, Behbahani:2011it}.  In the absence of a continuous shift symmetry, scale invariance may be significantly broken by sinusoidal features in power spectrum\footnote{There is also the interesting possibility that the continuous symmetry is restored in the infrared (IR), despite being broken explicitly \cite{Green:2009ds, LopezNacir:2011kk, Nacir:2012rm}.  These examples involve dissipative effects but give rise to scale invariant correlation functions.}.  In general, one also expects large deviations from scale invariance in other N-point correlation functions.  Nevertheless, it has been shown in the context of single field inflation that the dominant signal of these models is the oscillating feature induced in the power spectrum \cite{Behbahani:2011it}.

Ongoing observations will constrain (and possibly detect) correlation functions beyond the power spectrum, including the bispectrum and trispectrum.  When discussing these signatures, one typically assumes that all the N-point correlation functions are scale invariant at the same percent level accuracy as the power spectrum.  This assumption is well motivated by the shift symmetry that explains the scale invariance of the power spectrum.  If we were to break the shift symmetry by the interactions of the inflaton, one would expect radiative corrections to induce the same level of breaking in the power spectrum.  

The question we will investigate is whether violations of scale invariance can, in principle, be dominated by non-Gaussian correlation functions.  We will address this question by constructing models where the power spectrum is protected from large radiative corrections by collective symmetry breaking\footnote{Collective symmetry breaking is well-known for its use in little-Higgs model building \cite{ArkaniHamed:2001nc, ArkaniHamed:2002qy}.  It has also been applied to inflation to explain the absence of large corrections to the inflaton mass \cite{Kaplan:2003aj,ArkaniHamed:2003mz}.}.  This can be achieved by enlarging the symmetry group of the action such that scale invariance is protected by several, independent symmetries.  An individual interaction may break some of these symmetries, but scale invariance is maintained as long as a subgroup is preserved. Therefore, any scale dependent correlation function must be proportional to the product of all of the couplings necessary to completely break the symmetry.  As a result, these effects will occur at the same order in perturbation theory for any N-point function, thus removing the preference for the signal to be dominated by the power spectrum.  

In order to realize collective symmetry breaking, we introduce additional fields that transform linearly (by a phase rotation) in association with the shift of the inflaton.  Interactions of these fields will explicitly break the continuous shift symmetry, but they will leave an unbroken discrete symmetry.  As a result, the models we construct will realize a generalization of resonant non-Gaussianity \cite{Chen:2006xjb, Bean:2008na, Chen:2008wn, Flauger:2010ja, Behbahani:2011it}.  These additional fields contribute to the curvature perturbation through interactions that convert the isocurvature fluctuation into an adiabatic fluctuation, during inflation.  We discuss two classes of models: one where the conversion is perturbative and is described by quasi-single field inflation (QSFI) \cite{Chen:2009zp} and the other where the mixing is strong and is described by a single effective propagating mode.  In the two examples we present, there are regions of parameter space where the signal is largest in the bispectrum.

The organization of the paper is as follows:  In Section 2, we will review resonant non-Gaussianity and its signatures. In Section 3, we show how collective symmetry breaking can reduce the resonant signal in the power spectrum, with emphasis on the case of QSFI.  In Section 4, we compute the signals of the perturbative, QSFI example for both the power spectrum and bispectrum.  In Section 5, we present an example where the mixing with additional fields is strong and compute its signatures.  Section 6 contains a further discussion of the results.  We review the relationship between shift symmetries and scale invariance in the Appendix.

\section{Resonant Non-Gaussianity and Radiative Corrections}
In this section, we briefly review resonant non-gaussianity \cite{Flauger:2010ja} and its origin from a discrete shift symmetry in the effective field theory of inflation,  following \cite{Behbahani:2011it}.  Resonant non-gaussianity occurs when an interaction involving the scalar metric fluctuation, $\zeta$,  is oscillatory in time (e.g. $H_{\rm int} \supset \mu \cos(\omega_\star t) \dot \zeta^3$).  When the frequency of this oscillation, $\omega_\star$, is large compared to the Hubble scale, $H$, the correlation function is determined at the time when the modes have energy $\omega \sim \omega_\star$.  The time dependence of the oscillation is translated into scale dependence of the correlation function through the time of resonance,  $a( t_{\rm res}) \sim k / \omega_\star$.  The precise relationship between scale invariance and time translations is reviewed in the appendix.

It will be useful to express these results in terms of the Effective Field Theory (EFT) of Inflation.   The action of the EFT must be diffeomorphism invariant; explicit functions of time may appear but will be accompanied by a goldstone boson, $\pi$, that transforms non-linearly under the symmetry \cite{Creminelli:2006xe, Cheung:2007st}.  We will work in the decoupling limit ($\dot H \to 0$ with $\Mp^2 \dot H$ fixed) such that the action is an explicit function of $t + \pi$ with $\zeta = -H \pi$.  The goldstone boson transforms as $\pi \to \pi -\xi^0(t,\x)$ under a general diffeomorphism $x^\mu \to x^\mu + \xi^\mu(t,\x)$ which leaves the action invariant.

The Lagrangian for $\pi$ is determined at lowest order by the requirement that we have an FRW solution to Einstein's equations with $\pi = 0$.  It was shown in \cite{Creminelli:2006xe, Cheung:2007st} that the action takes the form
\bea
\L &=&-\Mp^2\left(3H^2(t+\pi) + \dot H \right)+ \Mp^2 \dot{H}\partial_\mu (t+\pi) \partial^\mu (t+\pi)+ \tfrac{1}{2} M_{2}^4\left(t+\pi\right)\left[ \partial_\mu (t+\pi) \partial^\mu (t+\pi) + 1\right]^2 \nonumber \\&&\qquad+\tfrac{1}{3!}M_{3}^4\left(t+\pi\right)\left[ \partial_\mu (t+\pi) \partial^\mu (t+\pi) + 1\right]^3+...\ .
\eea
If the couplings $M^4_n(t+\pi)$ are time independent, then the action has a continuous shift symmetry under $\pi \to \pi +c$.  The shift symmetry can be broken to a discrete subgroup if any of the couplings are periodic in time.  For example, in simplest model of slow-roll inflation,  this can be achieved using
\beq\label{eq:Hsocillatin}
H(t)=H_{\rm sr}(t+\pi)+H_{\rm osc}(t+\pi) \sin(\omega_\star \left(t+\pi\right)) .
\eeq
In this case, the continuous shift symmetry is broken by $H_{\rm osc}$ to a discrete one $\pi(\vec x,t)\rightarrow \pi(\vec x,t)+2\pi/\omega_\star$.  The discrete shift symmetry allows for self interactions of $\pi$ that do not contain derivatives.  Up to slow-roll corrections, these take the form
\beq
\label{Ln_pi}
{\cal L}_n(\pi)\simeq-\frac{1}{n!} \Mp^{2} \left(H^{(n+1)}+3HH^{(n)}\right)
               \pi^n\,.
\eeq
Using the above interaction Lagrangian we can calculate the higher correlation functions
\beq
\label{R_N}
\Expect{\prod_{i=1}^n\R_{\mathbf{k}_i}}=  (2\pi)^3\delta^3\left(\sum_{i=1}^n \mathbf{k}_i\right) A_nB_n(k_i) ,
 \eeq
where
 \bea
 A_n &\equiv& (-)^{n+1} \frac{ \epsosc \sqrt{2\pi}}{4} \alpha ^{2n-7/2}\left(\frac{H^2}{4\epsilon\Mp^2}\right)^{n-1}\, , \\
B_n(k_i)&\equiv& \frac{1}{ K^{n-3} \prod_i k_i^2} \left[ \sin\left(\alpha \ln(K/k_\star)\right)+\frac1\alpha \cos\left(\alpha \ln (K/k_\star)\right) \sum_{j,i} \frac{k_i}{k_j} \right] \, .
\eea
We have defined $\alpha\equiv\omega_\star/H\gg 1$ and $\epsosc = \frac{\alpha H_{\rm osc}}{\epsilon H}$, with $k_\star$ as the pivot scale.  From the quadratic action, we find oscillatory imprints on power spectrum, 
\bea
  \label{2pt}
\Expect{\R_\mbf{k}\R_\mbf{{k'}}} = (2\pi)^3 \delta^{(3)}(\mbf{k}+\mbf{k'}) \frac{H^2(t_\star)}{4\epsilon(t_\star) \Mp^2 k^3} \left[ 1- \left(\frac{\pi}{2}\right)^{1/2}\epsosc\, \alpha^{1/2}\sin\left(\alpha\ln(2k/k_\star)\right)\right]\, .
  \eea
As it is shown in \cite{Behbahani:2011it}, most of the scale-dependent signal is in the two point function. 

At the classical level, we are free to write the action such that the only oscillatory coupling involves more than two powers of $\zeta$.  In this sense, one could imagine generating oscillating non-Gaussian correlation functions without altering the power spectrum.  However, when radiative corrections are included, there is no reason to expect this result to hold.
As a concrete example, we will consider the interaction $H_{\rm int} = \mu \cos(\omega_\star t) \dot \zeta^3$, which arises in the effective theory through the operator
\bea
H_{\rm int}& =& \frac{M_3^4 \cos\Big(\omega_\star (t+\pi)\Big)}{3!} [ \partial_\mu (t+\pi) \partial^\mu (t+\pi) + 1]^3 \\
&=& \frac{M_3^4\cos(\omega_\star t)}{3!}[ 8 \dot \pi^3 + 12 \partial_\mu \pi \partial^\mu \pi \dot \pi^2 +\ldots] - \frac{M_3^4\sin(\omega_\star t)}{3!} [ 8 \omega_\star \pi \dot \pi^3 + \ldots] \ .
\eea
Here, and throughout the paper we will assume $\alpha \equiv \omega_\star / H > 1$.  To compute the radiative corrections, we will follow the strategy of \cite{Senatore:2010jy}, where we will estimate the loop contributions using a hard cutoff at $\omega =\Lambda$.  This cutoff is taken to be near the strong coupling scale, $\Lambda^2 \lesssim \Lambda^2_{\rm strong} = 4 
\pi (\Mp^2 |\dot H|)^{\frac{3}{2}} / M_3^4$, as this is the scale where our EFT is breaking down.  Because the correlation function is determined at energies of order $\omega_\star$, we further require $\omega_\star < \Lambda$ for our calculations to be under control.

We are most interested in the radiative correlations to the quadratic action for $\pi$.  The largest contribution arises from two insertions of $H_{\rm int}$ that renormalize $\dot \pi^2$ via a 1-loop diagram, yielding
\beq
\delta L_{\rm eff.} \supset \frac{M_3^8}{(\Mp^2 |\dot H|)^2} \frac{\Lambda^4}{16\pi^2} \cos^2\Big(\omega_\star (t+\pi)\Big) (\partial_\mu (t+\pi) \partial^\mu(t+\pi) + 1)^2 \simeq \Mp^2 |\dot H| \cos^2(\omega_\star t) \dot \pi^2 + \ldots \ .
\eeq
We see that we have an order one correction to the kinetic term of $\pi$ that depends explicitly on time.

It is straightforward to estimate the ratio of the signal from the resonance in bispectrum to the signal in the power spectrum by ${\cal L}_3 / {\cal L}_2$.  This ratio should be evaluated at the resonant frequency, $\omega \sim \omega_\star$, giving us
\beq
\frac{(S/N)_3}{(S/N)_2} \sim \frac{{\cal L}_3}{{\cal L}_{\rm eff}} |_{\omega \sim \omega_\star} \sim \frac{M_3^4}{\Mp^2 |\dot H|} \dot \pi |_{\omega = \omega_\star} \sim \frac{\omega_\star^2}{\Lambda_{\rm strong}^2} < 1 \ .
\eeq
Here we have dropped the $\cos(\omega_\star t)$ since it is of order one (for our purposes).  The last inequality is essentially the result of \cite{Behbahani:2011it}, which states that the signal is dominated by the power spectrum when the effective field theory is weakly coupled.

\section{Collective Symmetry Breaking of Shift Symmetries}\label{sec:collective}

As we saw above, given a generic action with some time dependent coefficient, all operators will acquire time dependent couplings of comparable size through radiative corrections.  Now we would like to understand if there are circumstances where this is not the case.

Scale invariance is the result of a shift symmetry.  This suggests that we can use this symmetry to protect the effective action from large radiative corrections.  Here we will use the idea of collective symmetry breaking to ensure that corrections to the power spectrum arise at higher order in the couplings.  
  
Collective breaking of a shift symmetry arises most naturally in multi-field models of inflation.  We would like $\pi$ to shift $\pi\to \pi+c$ under more than one $U(1)$ such that different $U(1)$s are only distinguished by the transformation(s) of the additional field(s).  We can only measure the correlation functions of $\zeta / \pi$, so the transformations of these additional fields are inconsequential for scale invariance.  To break the shift symmetry, one will need to break all of these $U(1)$s and doing so may require several non-zero couplings.  As a result, scale dependence in the $\pi$ correlation functions must be proportional to all of these couplings, regardless of which $N$-point function we are computing.  

The most straightforward implementation of this idea arises when $\pi$ mixes with an additional field, $\sigma$.  For now, let us focus on the case where this mixing is perturbative, namely quasi-single field inflation (QSFI) \cite{Chen:2009zp}.  We still study the observational consequences of these models in Section \ref{sec:qsfi} and the case of strong mixing in Section \ref{sec:strong}.  For the remainder of the section, we will use QSFI as the primary example only for simplicity.

\begin{figure}[h!]
   \centering
       \includegraphics[scale =0.29]{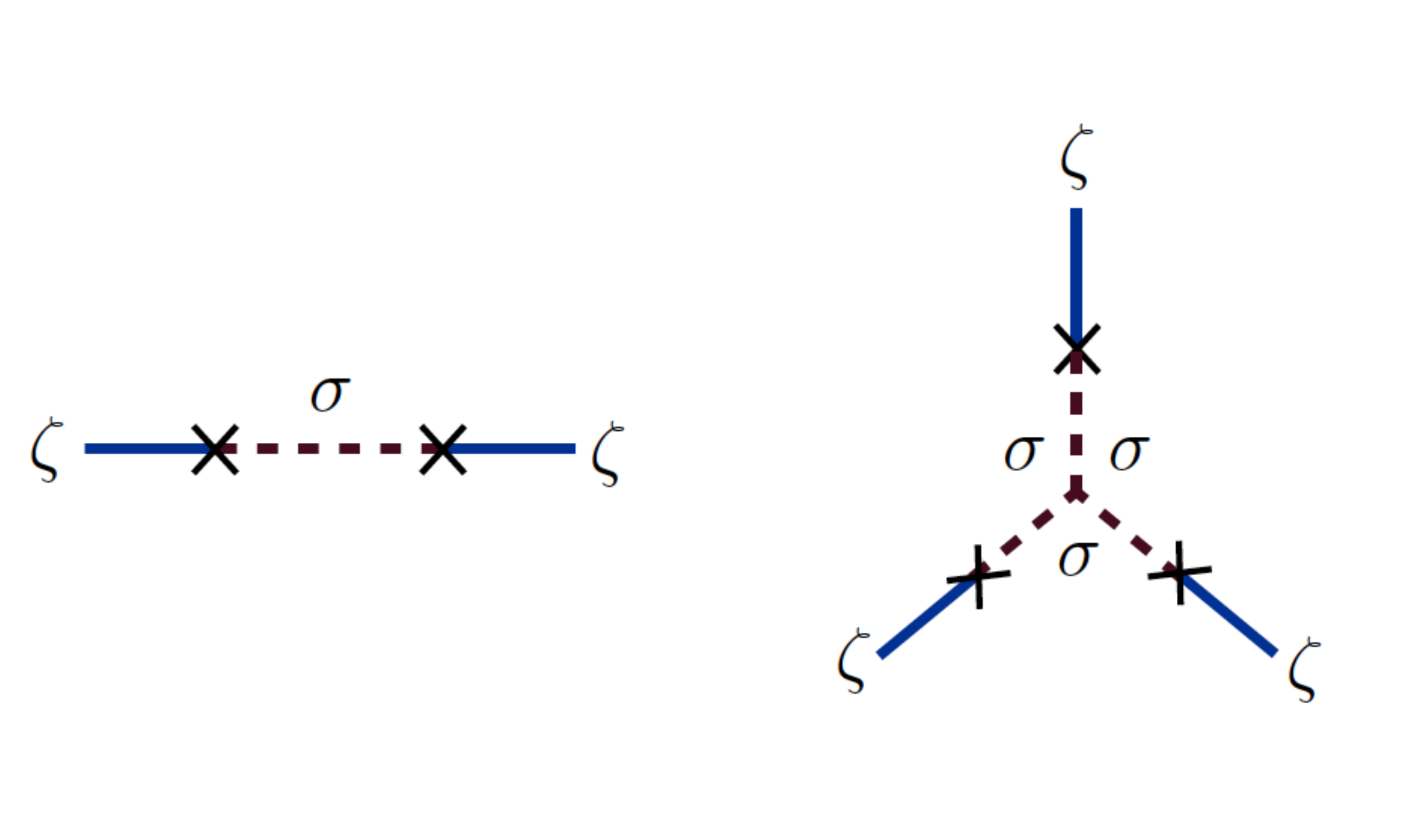}
   \caption{The leading perturbative contributions to the power spectrum (left) and bispectrum (right) of the curvature perturbation $\zeta$ that arise in quasi-single-field inflation.}
  \label{fig:Feynman}
\end{figure}

In QSFI, additional light fields contribute to the correlation functions through interactions during inflation (rather than through reheating / curvaton effects at late times).  To produce a signal, we require an additional field $\sigma$ with both self-interactions and a mixing interaction with $\pi$.  The non-gaussian correlation functions of $\sigma$ give rise to non-gaussian correlations of $\zeta$ through the conversion of $\sigma$ into $\zeta$ during inflation.  The contributions to the various $N$-point functions are illustrated via the appropriate Feynman diagram in Figure \ref{fig:Feynman}.

As a concrete example, we will consider the following Lagrangian
\bea
\label{equ:L1}
\L &=& \Mp^2 (3 H(t+\pi)^2 + \dot H(t+\pi)) + \Mp^2 \dot H \partial_\mu (t+\pi) \partial^\mu (t+\pi) -\partial_\mu \sigma \partial^\mu \bar \sigma  \nonumber \\ && \qquad 
\qquad +  \Mp |\dot H|^{1/2} \partial_\mu (t+\pi) \partial^\mu (\bar \beta\sigma+\beta \bar \sigma)[\partial_\nu (t+\pi) \partial^\nu (t+\pi)+1]  \\ &&  \qquad \qquad + \tfrac{1}{M^2} [ e^{i \omega_\star (t+\pi)} (\partial_\mu (t+\pi) \partial^\mu \sigma)^3 + {\rm h.c.}]\nonumber \ .
\eea
Here, $\sigma$ is a {\it complex} scalar field, while $\beta$ is a complex, dimensionless coupling and $\omega_\star$ is some fixed frequency.  For generic values of the couplings, this action explicitly breaks the shift symmetry $\pi \to \pi + c$ to a discrete symmetry $\pi \to \pi + 2 \pi n / \omega_\star$ with $n \in \mathbb{Z}$.  

In order to see that this breaking is ``collective", first consider the action when $\beta = 0$.  In this case, our Lagrangian possesses a exact symmetry, $U(1)_a$ where $\pi \to \pi + c$ and $\sigma \to e^{- i \omega_\star c / 3} \sigma$.  Alternatively, we may take $\beta \neq 0$ but send either $M \to \infty$ and / or $\omega_\star \to 0$.  In this limit, the action possesses an exact symmetry, $U(1)_b$ under $\pi \to \pi + c$ with $\sigma$ unchanged.  Therefore, the shift symmetry of $\pi$ (i.e. scale invariance) is only broken in the presence of both $\omega,\beta \neq 0$ and finite $M^2$.  

The consequences of these symmetries are most apparent if we promote $\beta$ to a background field such that $\beta \to  e^{- i \omega_\star c / 3} \beta$ under $U(1)_a$.  As a result, any correlation function of $\zeta$ containing the oscillation will be proportional to $\beta^3 e^{ i \omega_\star t_{\rm res}}$ such that the correlation function is formally invariant under the combined shift and dilatation.  Any oscillatory feature in the power spectrum or the bispectrum is therefore proportional to $\beta^3$.  This feature ensures that there is no enhancement of the signal in the power spectrum for small $\beta$.

In order to illustrate the implications of the collective breaking, let us consider the renormalization of $\dot \sigma^2$.  Expanding (\ref{equ:L1}) in powers of the fields, we find the Lagrangian
\bea
\L &=& -\tfrac{1}{2} \partial_\mu \pi_c \partial^\mu \pi_c -\partial_\mu \sigma \partial^\mu \bar \sigma+ \tfrac{1}{\sqrt{2}} \dot \pi_c ( \bar \beta \dot \sigma+ \beta \dot {\bar \sigma})\nonumber \\ && +   \tfrac{1}{M^2} [ e^{i \omega_\star t}\dot \sigma^3 + {\rm h.c.}]  +   \tfrac{\partial_\mu \pi_c}{\sqrt{2} \Mp |\dot H|^{1/2} M^2} [ 3 e^{i \omega_\star t}\dot \sigma^2 \partial^\mu \sigma + {\rm h.c.}]  \label{equ: L2} + \ldots \ ,
\eea
where $\pi = \sqrt{2}  \Mp |\dot H|^{1/2} \pi_c$ and the $\ldots$ include all terms that will give equal or subdominant contributions to the signal to noise.  Radiative corrections to the above action will generate the following operators
\beq\label{equ:radiative}
\L \supset \tfrac{3}{32 \pi^2} [\frac{\Lambda^4}{M^2 (\Mp |\dot H|^{1/2})} [ \beta e^{i \omega_\star (t+\pi)}\dot \sigma^2 + {\rm h.c.}] + \frac{ \Lambda^4}{M^2  (\Mp |\dot H|^{1/2})}  \dot \pi_c (\beta^2 e^{i \omega_\star (t+\pi)} \dot \sigma + {\rm h.c.})] \ .
\eeq
Both of these operators will contribute an oscillatory signal in the power spectrum.  However, unlike the previous section, these corrections are suppressed by $\beta$.  As explained above, by promoting $\beta$ to a background field, we see that operators containing fewer powers of $\sigma$ must contain additional powers of $\beta$, in order to maintain the invariance of the action under $U(1)_a$.

In writing the second term in \ref{equ:radiative}, we have dropped a tadpole for $\sigma$ of the form
\beq
\delta \L_{\rm tadpole} \sim - \omega_\star \frac{ \Lambda^4}{M^2} (\beta^2 e^{i \omega_\star (t+\pi)} \sigma + {\rm h.c.}) \ .
\eeq
Because we are studying the {\it fluctuations}, we simply cancel this tadpole by inserting the appropriate counter-term into the bare Lagrangian, $\delta \L_{\rm counter} \sim - \delta \L_{\rm tadpole}$.  One might worry that the presence of the tadpole suggests that $\sigma$ should have a potential with a natural scale set by the $\mu \sim [\beta^2  \omega_\star \frac{ \Lambda^4}{M^2 }]^{1/3}$.  Due to the suppression by $\beta^2$, the scale $\mu$ can be made to satisfy $\mu \ll H$ without suppressing the signals we will discuss below.

For the reader who is more familiar with the application of collective symmetry breaking to Higgs model building \cite{ArkaniHamed:2001nc, ArkaniHamed:2002qy}, the results in this section may appear slightly unusual.  Specifically, in a renormalizable theory like the standard model (and extensions thereof), collective symmetry breaking is expected to remove power-law divergences.  Here, we have suppressed radiative corrections by various small parameters, but we have not changed the power of the cutoff. Here our interactions are irrelevant and we generally expect power-law renomalization to arise.

In order to produce a complete model, we still require a UV completion above the strong coupling scale that respects the approximate symmetries discussed here.  It is also possible that such a UV completion would become important at a scale well below the strong coupling scale, in effect, lowering the effective cutoff $\Lambda$.  Lowering the cutoff would weaken the signal in the power spectrum and will be relevant to the generalizations we will discuss in Section \ref{sec:strong}.  For further discussion, see \cite{Baumann:2011su, Baumann:2011nk}.

\section{Perturbative Mixing : Resonance in Quasi-Single Field Inflation}\label{sec:qsfi}
In the previous section, we showed how collective symmetry breaking can weaken the size of radiative corrections to the quadratic action.  We are now in a position to compute the actual signatures produced by the action
\bea
\L &=& -\tfrac{1}{2} \partial_\mu \pi_c \partial^\mu \pi_c -\partial_\mu \sigma \partial^\mu \bar \sigma+ \tfrac{1}{\sqrt{2}} \dot \pi_c ( \bar \beta \dot \sigma+ \beta \dot {\bar \sigma})\nonumber \\ && +   \tfrac{1}{M^2} [ e^{i \omega_\star t}\dot \sigma^3 + {\rm h.c.}]  +   \tfrac{\partial_\mu \pi_c}{\sqrt{2} \Mp |\dot H|^{1/2} M^2} [ 3 e^{i \omega_\star t}\dot \sigma^2 \partial^\mu \sigma + {\rm h.c.}] + \L_{\rm rad.}\ ,
\eea
where $\L_{\rm rad.}$ are the radiative correlations to the effective action, including those in equation \ref{equ:radiative}. We first estimate the signal to noise generated in both the bispectrum and power spectrum, followed by a detailed calculation of the full momentum dependence and signal to noise.

\subsection{Estimating the Signal to Noise}

The oscillatory features in the power spectrum are controlled by the two operators in (\ref{equ:radiative}).  As both will contribute at the same order, let us pick one as an example, namely,
\beq
\L_{2} \supset \tfrac{3}{32 \pi^2}\frac{\Lambda^4}{ M^2 (\Mp |\dot H|^{1/2})} [ \beta e^{i \omega_\star t }\dot \sigma^2 + {\rm h.c.}] \ .
\eeq
Without loss of generality, we may take $\beta$ to be real, such that 
\beq
\L_{2}^{\rm osc.} = \tfrac{3}{32 \pi^2} \frac{\Lambda^4}{M^2 (\Mp |\dot H|^{1/2})} \beta \cos( \omega_\star t ) \dot \sigma_R^2 \ .
\eeq
where $\sigma_R = \tfrac{1}{\sqrt{2}} (\sigma +\bar \sigma)$.  We want to compare this signal with the leading contribution to the bispectrum, which arises from
\beq
\L_3 =  \tfrac{1}{\sqrt{2} M^2}\cos(\omega_\star t) \dot \sigma_R^3 \ .
\eeq
These operators contribute to the power spectrum and bispectrum of $\zeta$ respectively as a result of the mixing term $\beta \dot \pi_c \dot \sigma_R$, which convert the $\sigma_R$ fluctuations into fluctuations of $\zeta = - H \pi$.  

We will estimate the ratio of the signal to noise in the bispectrum, $(S/N)_3$, to signal-to-noise in the power spectrum $(S/ N)_2^{\rm osc.}$.  In a typical single field model, we would estimate this by $(\L_3)/ (\L_2^{\rm osc.})$ evaluated at horizon crossing, $\omega \sim H$.  There are two important subtleties that arise in these models.  First, in the case of resonance, the signal is generated at the resonant frequency $\omega = \omega_\star$ rather than horizon crossing.  Second, we want to estimate the signal in $\zeta$ correlation functions, not $\sigma$ correlation functions.  The mixing interaction is responsible for converting a $\sigma$ to a $\zeta$ with strength\footnote{Naively, the interaction strength is given by $\beta$.  The enhancement by $\alpha$ represents the extra time over which the mixing operates relative to conventional freeze-out.  This will be shown in detail in the next section.} $\beta \alpha$, where $\alpha \equiv \omega_\star / H$.  Therefore, our ratio of the signals-to-noise is given by
\beq\label{equ:signalesta}
\frac{(S/N)_3}{(S/ N)_2^{\rm osc.}} \sim \frac{\alpha^3\beta^3 \L_3|_{\omega = \omega_\star}}{\alpha^2 \beta^2 \L_2^{\rm osc.} |_{\omega = \omega_\star}} %\sim\frac{\tfrac{1 }{M^2} \beta^3 \cos(\omega_\star t)   \dot \sigma_R^3 }{\tfrac{\Lambda^4}{M^2 (\Mp |\dot H|^{1/2})} \beta^3 \cos( \omega_\star t ) \dot \sigma_R^2}
\sim \frac{\alpha (\Mp |\dot H|^{1/2}) \dot \sigma_R }{\Lambda^4 } \sim  \frac{ (\Mp |\dot H|^{1/2}) \omega_\star^3 }{\Lambda^4 H } \ .
\eeq
In order for this estimate to be reliable we require the resonance to occur below the strong coupling scale $\Lambda_\star \sim M$, or $\omega_\star < \Lambda_{\star}$.  Even if we take the UV cutoff to be the strong coupling scale $\Lambda \sim \Lambda_\star \sim M$, the non-gaussian signature dominates when
\beq
M \Big( \frac{M H }{(\Mp^2 |\dot H|)^{1/2}} \Big)^{1/3}< \omega_\star < M \ .
\eeq
Therefore, if $M^2 H^2  < \Mp^2 |\dot H|$ (or $M^2 < \epsilon \Mp^2 $) there is a range of $\omega_\star$ where the resonant signal is mostly non-gaussian.  In general $\Lambda < \Lambda_\star$, which would increase the range of parameters where the bispectrum gives the largest signal.

\subsection{Calculating the Bispectrum}

In this section we will present calculations of the bispectrum and power spectrum.   The calculations are nearly identical, therefore we will only show the bispectrum calculation in detail.

Cosmological correlation functions are given as equal time correlation function in the in-in formalism.  Following the standard presentation (see \cite{Weinberg:2005vy} for details), we compute correlations function in the interaction picture as
\beq
\langle \bar T \exp[i \int^{\tau_0}_{-\infty(1+i\epsilon)}  H_{\rm int}(\tau)d\tau ]\Big( \zeta_{\rm int}(\k_1, \tau_0) .. \zeta_{\rm int}(\k_n, \tau_0)  \Big)  T \exp[ -i \int^{\tau_0}_{-\infty(1-i\epsilon)}  H_{\rm int}(\tau) d\tau ] \rangle \ ,
\eeq
where we are working in conformal time $\tau$.  In practice, it is easier to perform this calculation after Wick rotation.  From the $i \epsilon$ prescription, it should be clear that we must rotate the time and anti-time ordered exponentials in the opposite way such that we are calculating
\beq
\langle \bar T \exp[i \int^{\tau_0}_{-i\infty +\tau_0}  H_{\rm int}(\tau)d\tau ]\Big( \zeta_{\rm int}(\k_1, \tau_0) .. \zeta_{\rm int}(\k_n, \tau_0)  \Big)  T \exp[ -i \int^{\tau_0}_{i \infty +\tau_0}  H_{\rm int}(\tau) d\tau ] \rangle \ .
\eeq
However, this is nothing other than the anti-time ordered correlation function in Euclidean time
\bea
&&\langle \bar T \Big( \zeta_{\rm int}(\k_1, \tau_0) .. \zeta_{\rm int}(\k_n, \tau_0) \exp[ i \int^{ i \infty+\tau_0}_{-i \infty +\tau_0}  H_{\rm int}(\tau) d\tau ] \Big)   \rangle \\
&& = \langle \bar T \Big( \zeta_{\rm int}(\k_1, \tau_0) .. \zeta_{\rm int}(\k_n, \tau_0) \exp[ - \int^{ \infty- i\tau_0}_{- \infty +\tau_0}  H_{\rm int}( - i\tau_E) d\tau_E ] \Big)   \rangle \ .
\eea
where $\tau = i \tau_E$.  It is now straightforward to calculate the bispectrum\footnote{We thank Kendrick Smith for this suggestion.}, provided we use the anti-time ordered, Euclidean Green's function
\beq
 \langle \bar T\Big(\sigma(\tau_E, \k_1) \sigma(\tau'_E, \k_2 ) \Big) \rangle  = \frac{H^2}{2 k_1^3} (1 + k_1 |\tau_E - \tau'_E| - k_1^2 \tau_E \tau'_E) e^{- k_1|\tau_E - \tau'_E| }  (2\pi)^2 \delta(\k_1 + \k_2)\ . 
\eeq 
The Green's functions for $\zeta$ are the same, up to the overall normalization which we define such that  $\zeta(\tau \to 0) = \frac{\Delta_\zeta}{k^{3/2}} $ with $\Delta_{\zeta} = 2.2  \times 10^{-4}$ \cite{Komatsu:2010fb}.  

We will compute the bispectrum, setting $\tau_0 = 0$ for simplicity and writing the Euclidean conformal time $\tau_E \to \tau$.  Our interaction Hamiltonian is given by
\beq
H_{\rm int} = (H \tau)^4 [ \beta \Mp |\dot H|^{1/2} \dot  \pi \dot \sigma + \frac{1}{\sqrt{2} M^2} \cos(\omega_\star t) \dot \sigma^3 ]  \ ,
\eeq
where $t =  -H^{-1}[ \log(\H \tau) - \tfrac{\pi}{2}]$ in de Sitter ($\dot H = 0$) and where have dropped the index on $\sigma_R\to \sigma$.  The bispectrum is given by
\beq
{\cal I} = \langle \zeta_{\k_1} \zeta_{\k_2} \zeta_{\k_3}(0) \int_{-\infty}^{\infty} d\tau_4  \prod_{i=1}^{3} [ \int_{-\infty}^{\infty} d\tau_i \frac{\beta \Mp |\dot H|^{1/2}}{(H \tau_i)^4} \dot  \pi_{-\k_i} \dot \sigma_{\k_i} (\tau_i) ] [ \tfrac{3!}{ \sqrt{2}} \frac{\cos(\omega_\star t_4)}{(H\tau_4)^4 M^2}  \dot \sigma_{-\k_1}  \dot \sigma_{-\k_2}  \dot \sigma_{-\k_3}(\tau_4)] \rangle
\eeq
Here we have implicitly contracted the fields, integrated over the momentum and dropped the overall momentum conserving delta function $(2\pi)^3 \delta(\k_1+\k_2 +\k_3)$.  Using the anti-time ordered Green's functions, this becomes
\bea
{\cal I} &=& \tfrac{3!}{ 2^5} (\frac{\Delta_\zeta^3}{k_1^{3/2} k_2^{3/2} k_3^{3/2}} ) \int_{-\infty}^{\infty} d\tau_4 \frac{H^2 \tau_4^2}{M^2} \cos(\alpha \log(H \tau_4) -i \alpha\pi/2) \prod_{i=1}^{3} \int^{\infty}_{-\infty} d\tau_i (\beta k_i e^{-k_i |\tau_i| - k_i |\tau_i -\tau_4|}) \nonumber \\
&=& \tfrac{3!}{ 2^5} ( \frac{ \Delta_\zeta^3}{k_1 k_2 k_3} \frac{\beta^3H^2}{M^2}) \int d\tau_4 \tau_4^2  \cos( \alpha \log(H \tau_4)- i \alpha\pi/2)  e^{- K |\tau_4| } \prod_{i=1}^3(1+ k_i |\tau_4|) \label{equ:fullbispectrum}\ .
\eea
This integral is straightforward to evaluate in general, although we are primarily interested in the leading behavior in $\alpha \gg 1$.  Because the dominant contribution to the integral is from $\tau \sim \alpha / K$, the leading contribution is from
\bea
{\cal I} &\simeq&  \tfrac{3!}{ 2^5} (\Delta_\zeta^3 \frac{\beta^3H^2}{M^2}) \int_{-\infty}^\infty d\tau_4 |\tau_4|^5 e^{- K |\tau_4| }\cos( \alpha \log(H \tau_4)-i \alpha\pi/2)  \\
&=&  \tfrac{3!}{2^5} \frac{ \Delta_\zeta^3}{K^6}\frac{\beta^3H^2}{M^2}  [ \tfrac{1}{2}( \Gamma(6+ i \alpha) K^{-i \alpha} + \Gamma(6- i \alpha) K^{+i \alpha}) e^{\tfrac{\pi}{2 }\alpha }\\&&\qquad \qquad +\tfrac{1}{2} (\Gamma(6- i \alpha) K^{+i \alpha} + \Gamma(6+ i \alpha) K^{-i \alpha})e^{-\tfrac{\pi}{2}\alpha }] \ .
\eea
Using Sterling's approximation, up to a $K$-independent phase, we find
\beq\label{equ:bispectruma}
\langle \zeta_{\k_1} \zeta_{\k_2} \zeta_{\k_3} \rangle \simeq  \tfrac{3!}{2^5} ( \frac{ \Delta_\zeta^3}{K^6} \frac{\beta^3H^2}{M^2}) \sqrt{2 \pi}  \alpha^{11/2} \cos(\alpha \log K/ k_\star) (2\pi)^3 \delta(\k_1+\k_2+\k_3) \ .
\eeq

\begin{figure}[h!]
   \centering
       \includegraphics[scale =0.9]{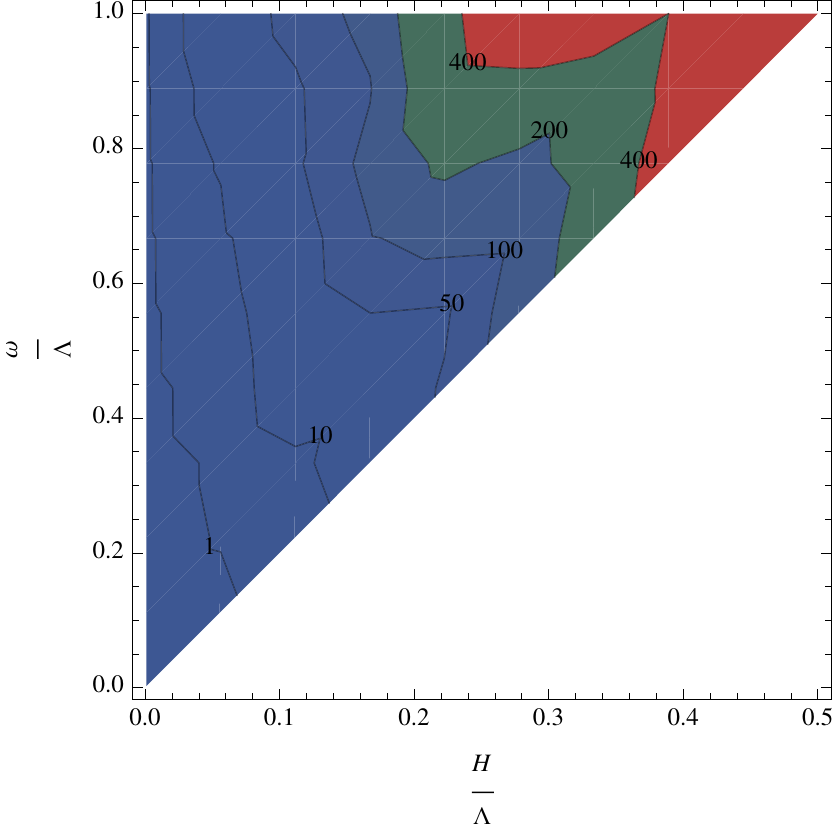}
   \caption{Contour Plot for $\frac{\left(\frac{S}{N}\left(\Expect{\zeta^3}\right)\right)}{\left(\frac{S}{N}\left(\delta\Expect{\zeta^2}\right)\right)}$. To ensure reliable results for the small values of $\alpha$, the signal to noise includes the full $\alpha$ dependence of the bispectrum given by equation \ref{equ:fullbispectrum} (and similarly for the power spectrum).   We have enforced $\alpha > 2$ to ensure that the our calculations and the theory are under control.  }
  \label{fig:SN}
\end{figure}

The power spectrum calculation is essentially identical, except that now
\beq
H_{\rm int} = (H \tau)^{-4} [ \beta \Mp |\dot H|^{1/2} \dot  \pi \dot \sigma + \gamma \cos(\omega_\star t) \dot \sigma^2 ]  \ ,
\eeq
where $\gamma  \simeq \tfrac{3}{32 \pi^2}\frac{\Lambda^4}{ M^2 (\Mp |\dot H|^{1/2})} \beta$ when it is generated from radiative corrections.  Following the steps in the previous section, we find for $\alpha \gg 1$
\beq\label{equ:powera}
\langle \zeta_{\k_1} \zeta_{\k_2} \rangle \simeq ( \frac{ \Delta_\zeta^2}{4 k_1^3} \beta^2 \gamma ) \sqrt{2\pi} (2 )^{-3} \alpha^{5/2} \cos(\alpha \log k_1/ k_\star) (2\pi)^3 \delta(\k_1+\k_2) \ .
\eeq
Comparing equations \ref{equ:bispectruma} and \ref{equ:powera}, we find good agreement with our parametric estimate in equation \ref{equ:signalesta}.
\vskip 8pt
It is straightforward to calculate the signal to noise, following the appendix of \cite{Behbahani:2011it}, including all numerical factors and the full momentum dependence.  To get an accurate estimate, we compute the bispectrum at all orders in $\alpha$ using equation \ref{equ:fullbispectrum}.  
This ratio is shown in Figure \ref{fig:SN} in terms of $\tfrac{H}{\Lambda}$ and $\tfrac{\omega_\star}{\Lambda}$.  

The signal to noise for much of  the parameter space is surprisingly small given the parametric estimates in the previous subsection.  This is due to a small numerical coefficient in $\left(\frac{S}{N}\left(\Expect{\zeta^3}\right)\right)$.  The suppression is most easily understood by considering the bispectrum \ref{equ:bispectruma} in the equilateral configuration $k_1 =k_2 =k_3$.  Because $K = 3 k_1$ in this configuration, the amplitude is suppressed by $3^{-6} \sim 10^{-3}$.  For this reason, the signal to noise ratio is on the order of $10-100$, rather than the $10^{4-5}$ that we expected.  Nevertheless, the bispectrum dominates over an appreciable range of parameters where the theory is under control.

\section{Strong Mixing : Resonance with a Linear Kinetic Term}\label{sec:strong}

In the previous sections, we looked at one particular realization of the collective breaking.  In this section, we will discuss a related example to show that the results were not particularly sensitive to being in the QSFI regime.

We will change the model from the previous section by altering the form of the mixing term.  The Lagrangian we will consider is instead
\bea
 \label{equ:L2}
 \L &=& \Mp^2 (3 H(t+\pi)^2 + \dot H(t+\pi)) + \Mp^2 \dot H \partial_\mu (t+\pi) \partial^\mu (t+\pi) -\partial_\mu \sigma \partial^\mu \bar \sigma \\ &&\nonumber \qquad 
\qquad +  \Mp |\dot H|^{1/2} (\bar \rho \sigma+\rho \bar \sigma)[\partial_\mu (t+\pi) \partial^\mu (t+\pi)+1]  \\ &&  \qquad \qquad + \tfrac{1}{M^2} [ e^{i \omega_\star (t+\pi)} (\partial_\mu (t+\pi) \partial^\mu \sigma)^3 + {\rm h.c.}]\nonumber \ .
\eea
Here $\rho$ is a complex coupling with dimensions of energy.  The only change to the above action from equation \ref{equ:L1} is in the second line. Keeping only the necessary operators and canonically normalizing $\pi \sqrt{2 \Mp^2 |\dot H|} = \pi_c$, we arrive at
\beq
{\cal L} =\tfrac{1}{2} \partial_\mu \pi_c \partial^\mu \pi_c + \partial_\mu \sigma \partial^\mu \bar \sigma + \tfrac{1}{\sqrt{2}} \dot \pi_c ( \bar \rho\sigma +  \rho \bar \sigma )  + \frac{1}{M^2} [e^{  i \omega_* t} \dot \sigma^3 + e^{-i \omega_* t} \dot{\bar{\sigma}}^{3}] + \ldots \ .
\eeq
When $\rho \ll H$, we expect the results to be identical to the previous section with $\beta \ll 1$ after replacing $\beta \to \rho / H$. Here, we will be interested in the regime where $\rho \gg H$. 

The dynamics of this model depend significantly on the energy scale \cite{Baumann:2011su}.  At high energies, $\omega \gg \rho$, this is a two field model with $\pi$ and $\sigma$ acting as the independent degrees of freedom.  When $\omega \ll \rho$, the mixing term dominates: $\rho \dot \pi \sigma \sim \rho \omega \pi \sigma \gg \dot \pi^2 \sim \omega^2 \pi^2$.  At these energies, $\sigma$ is not an independent degree of freedom, but rather $\rho \sigma \sim P_{\pi}$ is the canonical momentum of $\pi_c$.  We will be interested in $\omega_\star < \rho$ such that we are effectively dealing with a single field at resonance.  In this case,  $\rho \dot \pi_c \sigma_R$ is effectively the kinetic term for both $\pi$ and $\sigma$ and we may drop $\dot \pi_c^2$ and $\dot\sigma^2$ as they become irrelevant operators.  Therefore, the effective action is given by
\beq
{\cal L} =\rho \dot \pi_c \sigma_R -\tfrac{1}{2} \partial_i \pi_c \partial^i \pi_c - \tfrac{1}{2} \partial_i \sigma_R \partial^i\sigma_R  + \frac{1}{\sqrt{2} M^2} \cos(\omega_\star t) \dot \sigma_R^3  + \ldots \ .
\eeq
where we have taken $\rho$ to be real and $\sigma_R$ is the real part of $\sigma$.  The imaginary part of $\sigma$ will not play role so we will again drop the subscript, $\sigma_R \to \sigma$.  

The quadratic equations of motion are given by
\bea
\rho \dot \pi - \frac{k^2}{a^2} \sigma & =& 0 \ ,  \\
-\rho \dot \sigma + 3 H \sigma_r  - \frac{k^2}{a^2} \pi & =& 0 \ .
\eea
The first equation is just a constraint that requires $\sigma = a^2\rho \dot \pi_c / k^2$, so we see again that $\sigma$ is not an independent degree of freedom.  Plugging this back into the equations, one can solve for the the mode function of $\pi_c$ and are given by \cite{Baumann:2011su}
\beq
u_\k (\tau) = \tau^2 H^2 \sqrt{\frac{\pi}{8}} \frac{k}{\rho} (-\tau)^{1/2} H_{5/4}^{(1)} (\tfrac{1}{2} \tfrac{H}{\rho} (k \tau)^2 ) \ ,
\eeq
such that
\beq \label{equ:modes}
\pi_c(\k,\tau) = \hat a_\k u_\k + \hat a^{\dagger}_\k u^*_\k \ .
\eeq
In terms of the parameters of the model, the power spectrum (for $M \to \infty$) is given by
\beq\label{equ:power}
k^3 \langle \zeta^2 \rangle' \equiv \Delta_{\zeta}^2  = \frac{2 \Gamma(\tfrac{5}{4})^2}{\sqrt{\pi}} \frac{H^4}{\Mp^2 |\dot H|} \Big(\frac{\rho}{H} \Big)^{1/2} \sim \tfrac{1}{2} \frac{H^4}{\Mp^2 |\dot H|} \Big(\frac{\rho}{H} \Big)^{1/2} \ ,
\eeq
where $'$ indicates we have removed the $(2 \pi)^3 \delta(\k+\k')$.

The radiative corrections discussed in Section \ref{sec:collective} are identical provided we substitute $\beta \to \rho / \Lambda$.  However, the model is not free from dangerous correction, even when the shift symmetry for $\pi$ is exact.  This is because the action no longer possesses a shift symmetry for $\sigma$ and we must worry that a large mass $m^2 \sigma^2$ will be generated.  We require $m \ll \rho$ for the above dynamics to hold.  Fortunately, due to the modified dynamics at low energies, this mass can be made sufficiently small without modifying the model below the scale $\rho$ \cite{Baumann:2011nk}.

\subsection{Estimating the Signal to Noise}

Before doing a detailed calculation, we will estimate the signal in the bispectrum compared to the power spectrum.  The power spectrum is scale invariant at leading order, but receives contributions from radiative corrections.  The largest effect comes from the renormalization of the operator
\beq
\L_{2}^{\rm osc.} = \tfrac{3}{32\pi^2}\frac{\rho \Lambda^3}{\Mp |\dot H|^{1/2} M^2} \cos(\omega_\star t) \dot \sigma^2 \ .
\eeq
The largest contribution to the bispectrum will be generated by $\L_3 = \frac{1}{\sqrt{2}M^2} \cos(\omega_\star t) \dot \sigma^3$.  As before, we can estimate the ratio of the signal to noise to the bispectrum to the power spectrum as 
\bea
\frac{(S/N)_3}{  (S/N)_2^{\rm osc.}} &\sim& \frac{{\cal L}_3}{\L_{2}^{\rm osc.} }|_{\omega = \omega_\star} \sim \frac{ \tfrac{1}{\sqrt{2} M^2} \dot \sigma^3}{ \tfrac{\rho \Lambda^3}{\Mp \dot H^{1/2} M^2} \dot \sigma^2}|_{\omega = \omega_\star} \sim \frac{\Mp \dot H^{1/2}}{\Lambda^3} \frac{\dot \sigma}{\rho}|_{\omega= \omega_\star} \\
&\sim& \frac{\Mp \dot H^{1/2}}{\Lambda^3} \frac{\omega_\star^2 \pi_c}{(k^2/a^2)} \ .
\eea
To complete the estimate, we need to understand the behavior of $\pi_c$.  Because of the non-relativistic kinetic term, this is not really the ``canonically normalized" field in the sense that its mode functions are not purely functions of energy and momentum.  Instead, $\tilde \pi_c = \pi_c \rho^{1/2}$ is the canonical field, but is has units $ [\tilde \pi_c ] = [\tfrac{k}{a}]^{3/2}$.  Using this observation, with $\omega = k^2/ (a^2 \rho)$, we find
\bea
\frac{(S/N)_3}{  (S/N)_2^{\rm osc.}} &\sim&\frac{\Mp \dot H^{1/2}}{\Lambda^2} \frac{\omega^2_\star}{\rho^{1/2} (\tfrac{k}{a})^{1/2}} |_{\omega =\omega_\star} \sim \frac{\Mp \dot H^{1/2}}{\Lambda^2}\frac{\rho}{\Lambda} \Big( \frac{\omega_\star}{\rho} \Big)^{7/4}\ .
\eea
The bispectrum will produce the largest signal provided that
\beq
\rho \Big(\frac{\Lambda^2}{\Mp \dot H^{1/2}}\frac{\Lambda}{\rho}\Big)^{\tfrac{4}{7}}<\omega_\star<\rho \label{equ:range2}
\eeq
The upper bound $\omega_\star < \rho < \Lambda$ is an assumption of the setup.  Therefore, a necessary (but not sufficient) condition for satisfying (\ref{equ:range2}) is that $\Lambda^2 < \Mp |\dot H|^{1/2}$.

\subsection{Calculating the Bispectrum}

In order to determine the bispectrum to leading order, we must calculate
\beq\label{equ:bispec2}
\langle \zeta_{\k_1} \zeta_{\k_2} \zeta_{\k_3} \rangle = - i \frac{1}{\sqrt{2} M^2} \langle \zeta_{\k_1} \zeta_{\k_2} \zeta_{\k_3}(0) \int_{-\infty(1-i\epsilon)}^0 \frac{d \tau}{(-H \tau)^4} \dot \sigma^3 (\tau) \cos(\alpha \log|H \tau|) \rangle - {\rm c.c.} \ ,
\eeq
where $\dot \sigma^3(\tau) \equiv \int \frac{d^3 p_1}{(2\pi)^3} \frac{d^3 p_2}{(2\pi)^3} \frac{d^3 p_3}{(2\pi)^3} \dot \sigma_{\p_1}\dot \sigma_{\p_2}\dot \sigma_{\p_3} (2\pi)^3\delta^3(\p_1 +\p_2 +\p_3)$ and $\alpha \equiv \omega_\star / H$.  Because $\sigma = \rho a^2 \dot \pi_c / k^2$, we can determine $\dot \sigma$ as an operator by differentiating $\pi_c$, using (\ref{equ:modes}).  Therefore, we need to evaluate
\bea
\dot{ \tilde{ \sigma}}_\k =  \partial_t \Big( \frac{ \rho a^2 \dot u_\k}{ k^2} \Big) &=& (-H \tau) \partial_\tau  \Big( \frac{ \rho \partial_\tau u_\k}{ - H \tau k^2} \Big) \nonumber \\ 
&=&  \sqrt{\frac{\pi}{8}}  \frac{H^3 k}{\rho^2} \tau^{5/2} \Big( H k^2 \tau^2 H_{-3/4}^{(1)}(\tfrac{1}{2} \tfrac{H}{\rho} (k \tau)^2)+ 2 \rho H_{1/4}^{(1)}(\tfrac{1}{2} \tfrac{H}{\rho} (k \tau)^2) \Big) \label{equ:derivative} \ .
\eea
where $\sigma_\k = \hat a_\k \tilde \sigma_\k + \hat a^{\dagger}_\k \tilde \sigma_\k^*$.

The integral in (\ref{equ:bispec2}) is dominated by the saddle point at $\tau^2  =\tfrac{\alpha}{\tilde K^2} \tfrac{\rho}{H}$ with $\tilde K^2 = k_1^2 + k_2^2+k_3^2$.  Because we are interested in the behavior  when $\alpha \gg 1$, we can keep only the term with the most factors of $\tau$, as it gives the largest power of $\alpha$.  The leading contribution to (\ref{equ:bispec2}) is given by
\bea
\langle \zeta_{\k_1} \zeta_{\k_2} \zeta_{\k_3} \rangle' & \simeq &  \frac{ 3!}{\sqrt{2} M^2}\frac{H^8 (k_1 k_2 k_3)^{3/2}}{\rho^6} \frac{\pi^{3/2} \Delta_\zeta^3}{8 ^{3/2}}  \\ && \times 2 \, {\rm Re} \int_{-\infty(1-i\epsilon)}^0 \hskip -20pt d\tau \tau^{19/2} \cos(\alpha \log|H \tau|) \prod_{i = 1}^3
 H_{-3/4}^{(2)}(\tfrac{1}{2} \tfrac{H}{\rho} (k_i \tau)^2) \\
 & \simeq&  \frac{ 3! H^2}{\sqrt{2} M^2}\frac{H^{9/2} ( k_1 k_2 k_3)^{1/2}}{\rho^{9/2}} \frac{\Delta_\zeta^3}{8^{1/2}}  \\
 && \times 2 \, {\rm Re} \int_{-\infty(1-i\epsilon)}^0 \hskip -20pt d\tau \tau^{13/2} \cos(\alpha \log|H \tau|) 
 e^{-i\tfrac{1}{2} \tfrac{H}{\rho} (\tilde K \tau)^2}  \\
 &\simeq& \frac{ 3! H^2}{ M^2}\frac{ H^{3/4} ( k_1 k_2 k_3)^{1/2}}{2^{1/4}\rho^{3/4} \tilde K^{15/2} } \Delta_\zeta^3 2\, {\rm Re} \Big( e^{\tfrac{1}{4} \pi \alpha}  \tilde K^{ i \alpha} \Gamma[\frac{15}{4} - i \frac{\alpha}{2} ] \Big) \\
 &\simeq&\frac{ 3! H^2}{ M^2}\frac{H^{3/4} ( k_1 k_2 k_3)^{1/2}}{ 4 \rho^{3/4} \tilde K^{15/2} } \Delta_\zeta^3  \sqrt{\pi} \,\alpha^{13/4}\cos( \alpha \log\tilde K / k_\star ) \ ,
  \eea 
where we have absorbed a $k$-independent phase into the definition of the pivot scale $k_\star$.

The oscillatory signal in the power spectrum follows from a nearly identical calculation.  Following the same steps as above, we find
\bea
\langle \zeta_{\k_1} \zeta_{\k_2} \rangle' &\simeq& \tfrac{3}{32\pi^2} \frac{\Delta_\zeta^2}{k_1^3} \Big( \frac{\rho \Lambda^3}{\Mp |\dot H|^{1/2} M^2}   \Big) 2 {\rm Re} \Big( \frac{H}{4 \rho} e^{\tfrac{1}{4} \pi \alpha } k_1^{i \alpha} \Gamma[2 - i \tfrac{\alpha}{2}]  \Big) \\
&\simeq& \tfrac{3}{32\pi^2}  \frac{\Delta_\zeta^2}{k_1^3} \Big( \frac{ H \Lambda^3}{\Mp |\dot H|^{1/2} M^2} \sqrt{\pi} (\alpha)^{3/2} \cos(\alpha \log k_1 / k_\star ) \Big) \ .
\eea

\begin{figure}[h!]
   \centering
       \includegraphics[scale =0.9]{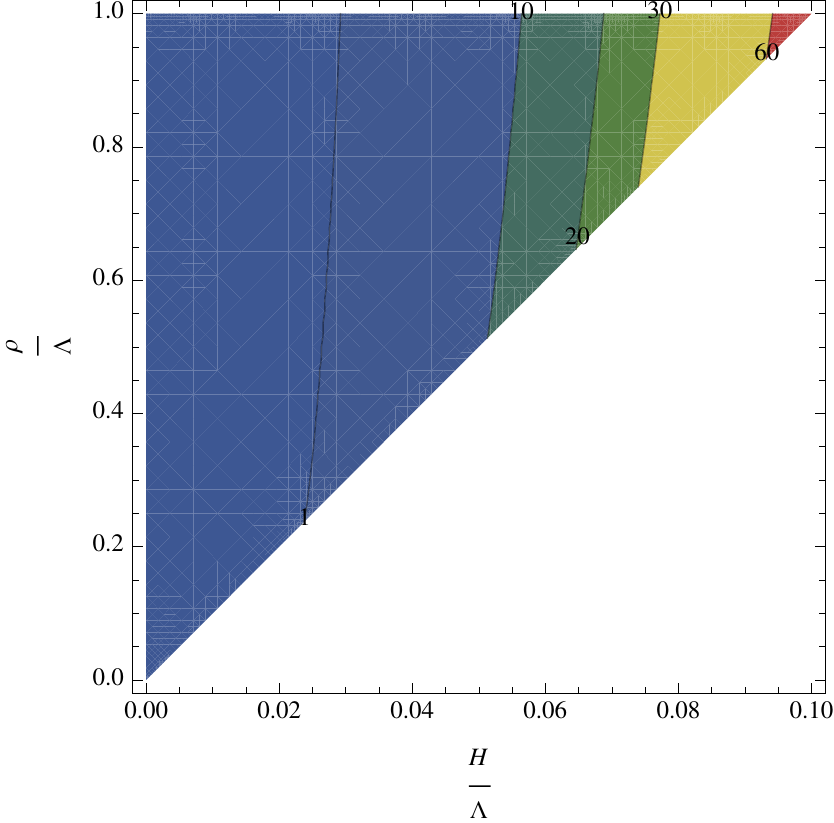}
   \caption{Contour Plot for  $\frac{\left(\frac{S}{N}\left(\Expect{\zeta^3}\right)\right)}{\left(\frac{S}{N}\left(\delta\Expect{\zeta^2}\right)\right)}$ with $\alpha \equiv \omega_\star / H =10$.  The signal to noise from the bispectrum and the power spectrum was calculated using the full $\alpha$ dependence from equations \ref{equ:bispec2} and \ref{equ:derivative}. }
  \label{fig:SN2}
\end{figure}

As in the previous sections, we can use the results of \cite{Behbahani:2011it} to give a more precise estimate of the signal-to-noise.  To achieve accurate results for moderate values of $\alpha$, we determine the full $\alpha$ dependence by including both terms from (\ref{equ:derivative}) when computing the bispectrum and power spectrum. The signal to noise ratio is shown in Figure \ref{fig:SN2} as a function of $\tfrac{\rho}{\Lambda}$ and $\tfrac{H}{\Lambda}$ for fixed $\alpha = 10$.  From our naive  estimates, we might have expected a smaller signal than what we found the QSFI example in the previous section.  Specifically, the signal in the bispectrum is only enhanced by $\alpha^{7/4}$ relative to the power spectrum  compared to $\alpha^3$ in QSFI; yet, the signal to noise estimate gives comparable results for physical values of $\alpha$. The main reason is that numerical suppression of the signal to noise is smaller than in the previous section.  We can again estimate the suppression by looking at the equilateral limit $k_1 = k_2 =k_3$. In this configuration, $\tilde K = \sqrt{k_1^2 + k_2^2 +k_3^2} \sim \sqrt{3} k_1$ and our bispectrum is suppressed by $3^{-15/4} \sim 10^{-2}$.  This is larger by a factor of ten from the QSFI example and explains why the results are comparable for $\alpha \sim{\cal O} (10)$.

\section{Discussion of results}

We studied the possibility that collective breaking of a shift symmetry could give rise to a large oscillatory signal in the bispectrum without introducing larger oscillations in the power spectrum.  We constructed two examples where the radiatively induced signal was suppressed relative to the bispectrum.  The common feature of these models is the presence of an additional scalar field that transforms by a phase in association with the shift of the inflaton. 

Scale invariance in the power spectrum and the bispectrum is broken in our models through oscillations in momentum space that arise as a result of resonance \cite{Flauger:2010ja, Behbahani:2011it}.  Due to the rapid oscillations, it is unclear that the signal to noise in the CMB is well approximated by the signal in the primordial correlation functions.  Given that the signal in the power spectrum is smaller than in the bispectrum by a factor of ten to a hundred, a more careful analysis of the signal in the CMB is required.

The broader motivation for this work was the question of whether non-gaussian correction functions can naturally contain generic, order one violations of scale invariance (i.e. including non-oscillatory shapes).  This is relevant to future observations of scale dependent bias \cite{Sefusatti:2012ye, Norena:2012yi} and $\mu$-distortion \cite{Pajer:2012vz}, where there is a degeneracy between the effects of massive fields \cite{Chen:2009zp,Baumann:2011nk} or excited states \cite{Ganc:2012ae, Agullo:2012cs} and a scale dependent amplitude for $f^{\rm local}_{\rm NL}$ \cite{Sefusatti:2009xu, Shandera:2010ei}.  It was crucial to our above constructions that we only broke the shift symmetry to a discrete subgroup.  Power law violation of scale invariance would require that the shift symmetry is completely broken.   The approach of collective symmetry breaking used in this paper would not protect such models from large radiative corrections.  It remains an open question whether large, power-law violations of scale invariance can arise in the bispectrum without fine-tuning.

\acknowledgments
We thank Leonardo Senatore for early collaboration.  We thank Daniel Baumann, Adam Brown, Raphael Flauger, Mehrdad Mirbabaei, and Kendrick Smith for helpful discussions. 
The research of S.R.B is supported by the DOE under grant numbers DOE grant DE-FG02-01ER-40676 and DE-FG02-01ER-40676.
The research of D.G.~is supported by the DOE under grant number DE-FG02-90ER40542 and the Martin A.~and Helen Chooljian Membership at the Institute for Advanced Study. 
\newpage

\appendix
\section{Scale Invariance and Shift Symmetries}

In this appendix, we will briefly review the relation between shift symmetries and scale invariance.  An FRW universe possesses a dilatation symmetry that arises from a large diffeomorphism.  This symmetry is realized non-linearly by the curvature perturbation, $\zeta$, and takes the form in co-moving coordinates \cite{Weinberg:2003sw} (see \cite{ Hinterbichler:2012nm, Senatore:2012wy, Assassi:2012zq} for further discussion)
\begin{align}
\x \ \to\ (1+\lambda) \x\  \qquad \zeta \ \to\ \zeta + \lambda ( 1 + \x \cdot \partial_\x\hskip 1pt \zeta ) \ ,
\end{align}
where $\lambda$ is an infinitesimal parameter.  Due to the presence of this symmetry, there must exist a solution $\zeta(x,t) = \zeta_0$, which is the limit of a finite momentum solution \cite{Weinberg:2003sw}. This ensures that correlation functions of long wavelength modes are time independent.  However, because the transformation acts non-linearly on $\zeta$, the correlation functions are not required to be scale invariant.  

Formally, the action for $\zeta$ is also invariant under the rescaling $x \to \lambda x$ and $a \to \lambda^{-1} a$.  However, the scale factor, $a(t)$, is itself a function of time and therefore this symmetry is broken by any time dependent coupling.  For example, in de Sitter, by writing $t = H^{-1} \log a$, we see that a time dependent coupling $\lambda(t) \to \tilde \lambda(a)$ breaks this symmetry.  On the other hand, if the action is invariant under $t \to t + c$, then our rescaling symmetry is also preserved.  Under the rescaling, $\zeta_k(t) \to \lambda^{-3} \zeta_k(t)$.  In addition, $\zeta_k$ must be time independent as $k \to 0$ which implies $\zeta_k (t) \to k^{-3}$ in this limit.  In other words, the correlation functions of $\zeta$ are scale invariant in the long wavelength limit, if the action possesses a time translation symmetry.  

The physics of freeze-out also makes this connection clear.  The correlation functions are typically determined by the couplings in the effective action for $\zeta$ at the time of horizon crossing, e.g. for slow roll when $k = a(t_\star) H(t_\star)$.  When we observe modes at different values of $k$, they were sensitive to couplings at different times $t_\star(k)$.  Hence, the time evolution of the couplings is translated into scale dependence through the freeze-out time $t_\star(k)$.  In resonant non-gaussianity, the role of freeze-out is played by the time when the mode crosses a resonance, $k = \omega_\star a(t_{\rm res.})$.  The general conclusion will be the same in both cases, couplings that are time dependent during inflation introduce scale dependence in late time observables through the freezing of the correlation functions at some fixed physical scale.

Although it may not seem clear from this presentation, this is closely related to having shift symmetry for the inflaton.  This can be most easily understood in the context of the effective theory of inflation.  During inflation, the time diffeomorphisms are broken by the inflaton background, and are only realized non-linearly in the theory of the fluctuations.  The time diffeomorphisms are maintained through a shift of the goldstone boson, $\pi$, provided that the action is a generic function of $t+\pi$.  As a result, if the action is time independent, then the symmetry $t \to t+c$ implies there is a symmetry under which $(t+\pi) \to (t +\pi) + c$.  This symmetry can also be interpreted as the statement that the action is invariant under a shift symmetry $\pi \to \pi+ c$.

\newpage
\begingroup\raggedright

\endgroup

\end{document}